\documentclass[journal, final, twoside, a4paper]{IEEEtran}
\usepackage{cite}
\usepackage{amsmath,amssymb,amsfonts}
\usepackage{algorithmic}
\usepackage{graphicx}
\usepackage{textcomp}
\usepackage{xcolor}
\usepackage{multirow}
\usepackage{url}

\usepackage[acronym]{glossaries}
\newacronym{sdo}{SDO}{Standardization Organization}
\newacronym{sno}{SNO}{Satellite Network Operator}
\newacronym{ns}{NS}{Network Slicing}
\newacronym{vnf}{VNF}{Virtual Network Function}
\newacronym{pnf}{PNF}{Physical Network Function}
\newacronym{nf}{NF}{Network Function}
\newacronym{st}{ST}{Satellite Terminal}
\newacronym{dvbs2}{DVB-S2}{Digital Video Broadcasting - Satellite - Second Generation}
\newacronym{nms}{NMS}{Network Management System}
\newacronym{snmp}{SNMP}{Simple Network Management Protocol}
\newacronym{sdn}{SDN}{Software Defined Network}
\newacronym{nfv}{NFV}{Network Function Virtualization}
\newacronym{miot}{MIoT}{Massive Internet of Things}
\newacronym{ims}{IMS}{IP Multimedia Subsystem}
\newacronym{leo}{LEO}{Low Earth Orbit}
\newacronym{meo}{MEO}{Middle Earth Orbit}
\newacronym{geo}{GEO}{Geostationary Orbit}
\newacronym{vhts}{VHTS}{Very High Throughput Satellite}
\newacronym{capex}{CAPEX}{Capital Expenditures}
\newacronym{opex}{OPEX}{Operating Expenses}
\newacronym{svno}{SVNO}{Satellite Virtual Network Operator}
\newacronym{satcom}{SatCom}{Satellite Communication}
\newacronym{ran}{RAN}{Radio Access Network}
\newacronym{cn}{CN}{Core Network}
\newacronym{tn}{TN}{Transport Network}
\newacronym{mvno}{MVNO}{Mobile Virtual Network Operator}
\newacronym{csc}{CSC}{Communication Service Customer}
\newacronym{csp}{CSP}{Communication Service Provider}
\newacronym{cs}{CS}{Communication Service}
\newacronym{ai}{AI}{Artificial Intelligence}
\newacronym{mno}{MNO}{Mobile Network Operator}
\newacronym{so}{SO}{Satellite Operator}
\newacronym{sp}{SP}{Service Provider}
\newacronym{bng}{BNG}{Broadband Network Gateway}
\newacronym{iaas}{IaaS}{Infrastructure as a Service}
\newacronym{ott}{OTT}{Over The Top}
\newacronym{api}{API}{Application Programming Interface}
\newacronym{sla}{SLA}{Service Level Agreement}
\newacronym{nss}{NSS}{Network Slice Subnet}
\newacronym{nssi}{NSSI}{Network Slice Subnet Instance}
\newacronym{qos}{QoS}{Quality of Service}
\newacronym{cos}{CoS}{Class of Services}
\newacronym{sba}{SBA}{Service Based Architecture}

\def\BibTeX{{\rm B\kern-.05em{\sc i\kern-.025em b}\kern-.08em
    T\kern-.1667em\lower.7ex\hbox{E}\kern-.125emX}}

\begin{document}

\title{An Extensible Network Slicing Framework for Satellite  Integration into 5G}


\author{
\IEEEauthorblockN{Youssouf Drif\IEEEauthorrefmark{1}\IEEEauthorrefmark{2}, Emmanuel Chaput\IEEEauthorrefmark{2}, 
Emmanuel Lavinal\IEEEauthorrefmark{2}, Pascal Berthou\IEEEauthorrefmark{3}, Boris Tiomela Jou\IEEEauthorrefmark{4}, \\Olivier Gremillet\IEEEauthorrefmark{1}, Fabrice Arnal\IEEEauthorrefmark{5}}\\
\IEEEauthorblockA{\IEEEauthorrefmark{1}IRT Saint-Exup\'ery, France; Email: youssouf.drif,olivier.gremillet@irt-saintexupery.com}\\
\IEEEauthorblockA{\IEEEauthorrefmark{2}IRIT, Toulouse, France; Email: emmanuel.chaput,emmanuel.lavinal@irit.fr}\\
\IEEEauthorblockA{\IEEEauthorrefmark{3}LAAS, Toulouse, France; Email: pascal.berthou@laas.fr}\\
\IEEEauthorblockA{\IEEEauthorrefmark{4}Airbus Defence and Space, Toulouse, France; Email: boris.tiomela-jou@airbus.com}\\
\IEEEauthorblockA{\IEEEauthorrefmark{5}Thales Alenia Space, Toulouse, France; Email: fabrice.arnal@thalesaleniaspace.com}
}

\maketitle


\section*{Context}

For the past decades, networks have evolved to increase their performances, their capacities, to reduce latencies and optimize their resource management in order to remain competitive and adapted to the market. Today, the way consumers use networks has changed and more heterogeneous services with their own requirements have emerged. This has led network operators to define the network slicing paradigm. 

Network slicing creates multiple partitions in the network, each partition can be dedicated to a particular service allowing vertical markets and multiple services with different requirements to run on top of a single infrastructure. Network slicing opens up new opportunities, aims at providing end-to-end services across multiple administrative network domains and is expected to facilitate the integration of the satellite into terrestrial networks.

To allow the flexibility level required by network slicing, satellite technologies have to evolve. Satcoms actors have therefore been working on improving satellite equipments.
On the one hand, researches conducted in projects such as H2020 VITAL\cite{vital_consortium_h2020_2015, ahmed_software_defined_2018, ferrus_sdn/nfv-enabled_2016,ferrus_towards_2017} focused on the application of \gls{sdn} and \gls{nfv} paradigms in satellite networks resulting in a flexible satellite network management.
On the other hand, projects such as H2020 SaT5G\cite{sat5g_consortium_satellite_2017} studied the integration of the satellite into 5G networks, initiating various ways of integration and identified connection points as well as possible hybrid terrestrial-satellite architectures and end-to-end network management system\cite{jou_architecture_2018}. Testbeds followed the theoretical analysis and ESA's current project SATis5\cite{noauthor_satis5_2018} is making it its core topic.

The work presented in the full-paper is the next step we propose to those initiatives. It focuses on the network slicing concept applied to satellite networks which, we believe, is a mandatory requirement for a full integrated satellite into 5G networks. We first start by describing the main challenges associated to the satellite slice definition. We then highlight a set of requirements for such a satellite slice. Based on those requirements, we construct and propose a complete Satellite Slice as a Service (S\textsuperscript{3}) framework which mutualizes the satellite infrastructure to provide a seamless integration into 5G networks.

\section*{Network slicing for satellite networks}

Applying the network slicing paradigm to the satellite is not straightforward, fundamental questions must be addressed.
We propose an in-depth analysis of the network slicing in 5G networks\cite{ts_23501, ts_28530, ts_38300} and more generally in the terrestrial networks\cite{afolabi_network_2018,ordonez-lucena_network_2017, open_networking_foundation_onf_applying_2016}, then we pinpoint similarities between 5G networks and satellite networks as well as differences, and finally we are giving our definition of network slicing for satellite.

A network slice is more than a simple network partition. 
It is firstly supposed to considerably increase the flexibility of the network management for a network operator and secondly be able to abstract the infrastructure complexity. This allows a tenant without network management knowledge to run services on top of the infrastructure.

Services running on top of a satellite slice could be satellite specific services as well as more generic ones\cite{jou_architecture_2018}. It could be for instance a satellite connectivity with enhanced services or an \gls{ott} service using the satellite connectivity.
An important feature of satellite slicing is to ease the creation procedure and management of \glspl{svno} for a \gls{sno}. Satellite slicing should offer the ability to change the level of flexibility that a \gls{svno} has on its slices from simplified \glspl{api} to complete management of the infrastructure. 
Moreover it should fit the 3GPP specifications on network slicing as the satellite could be integrated in various ways into 5G networks (from a simple Transport Network to a fully integrated 5G-satellite network as explained in SaT5G\cite{tiomela_jou_integrated_2018}).

From the aforementioned network slicing analysis and the associated challenges, we extract and summarize a set of slicing requirements such as slice network isolation, slice life-cycle management or slice orchestration. Additionally, we consider satellite network specific requirements, we then categorize all the requirements making them the key elements we rely on for the S\textsuperscript{3} framework design and validation.

\begin{figure*}[!ht]
    \centering
    \includegraphics[width=18cm]{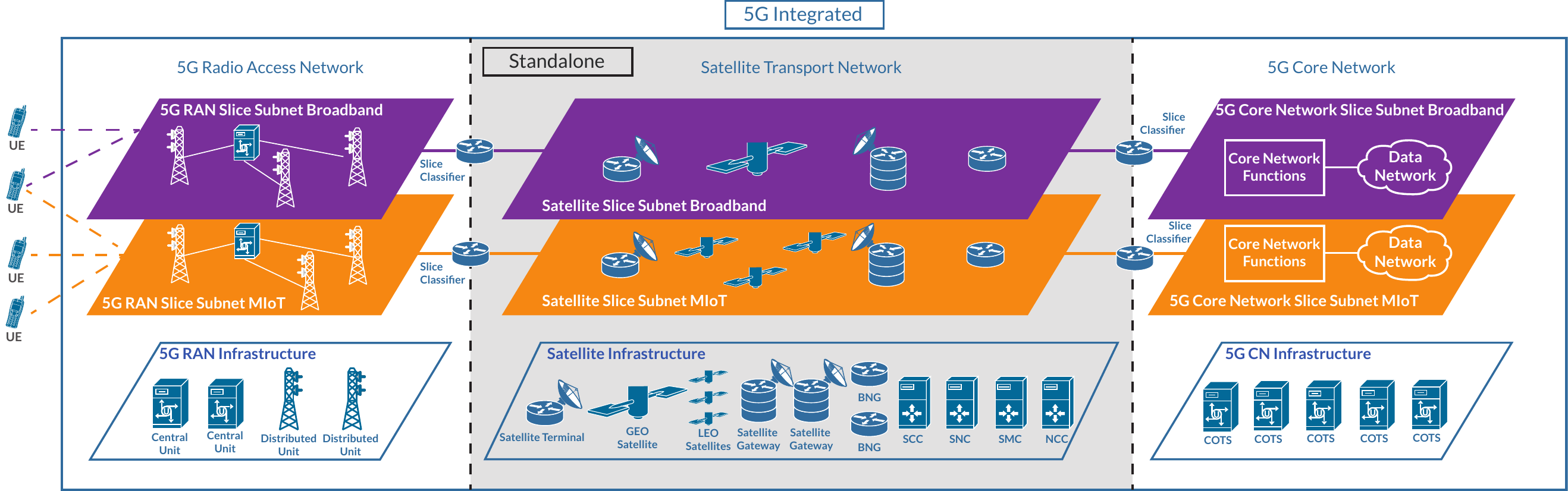}
    \caption{S\textsuperscript{3} 5G integrated mode and Standalone mode}
    \label{fig:framework_application_domain}
\end{figure*}

\section*{Satellite Slice as a Service (S\textsuperscript{3})}

Empowered by \gls{sdn} and \gls{nfv}, we build S\textsuperscript{3} which includes defining, modelling, deploying and orchestrating  multiple satellite slices on top of a mutualized satellite infrastructure.

As shown in Figure \ref{fig:framework_application_domain}, we consider the satellite as a \gls{tn} and model the satellite as a 3GPP \gls{nss}\cite{ts_28530}: the satellite is a network slice subnet providing all the necessary mechanisms to communicate and handle flows from 5G \gls{ran} and \gls{cn}.
For this integration, we proceed mainly as follows:

\begin{itemize}
    \item We define \glspl{api} between the 3GPP management system and the satellite management system (considered as the \gls{tn} manager in the 3GPP specification\cite{ts_28530}): through those \glspl{api}, the management systems will exchange information such as slice creation/modification requests, 5G \gls{qos} parameters or performance constraints. Those interfaces are the main exchange points between the 5G network and the satellite network control plane.
    \item We upgrade the satellite HUBs with \gls{sdn}/\gls{nfv}: the current satellite architecture can be challenging to support the dynamism inferred by 5G networks, therefore we push forward the work on the virtualization of the satellite gateway \cite{ahmed_software_defined_2018}. We now consider the satellite HUB as a pool of resources and propose to dynamically build the satellite gateways based upon slice requirements and resource availability. Inspired by the 5G \gls{sba}, we model the gateway as a composable succession of \glspl{nf}. The customization of the gateway allows to fully exploit the various satellite types such as LEO, MEO and GEO and to benefit from the current work done in 3GPP on 5G.
    \item We introduce Slice Classifiers: those classifiers are the components which stitch the different network slice subnets of the end-to-end slice as shown on Figure \ref{fig:framework_application_domain}. They answer the challenges related to the flow handling between the 5G \gls{ran}, 5G \gls{cn} and the satellite network slice subnet.
\end{itemize}


Furthermore, we generalize our approach of the network slicing concept for satellite networks making the framework usable in other context than in 5G. This feature makes S\textsuperscript{3} extensible by design: it can be used to establish and orchestrate end-to-end slices through the satellite and the 5G terrestrial \gls{ran} and \gls{cn} as describe above, or be used in a standalone version.
In its standalone version, the framework is used to deploy and orchestrate ``pure satellite slices'': the satellite is not a 5G network slice subnet anymore and can be considered as an end-to-end slice offering customizable \glspl{api}. 

In our future work we plan to implement S\textsuperscript{3} on a real testbed first and then experiment multiple use cases in order to demonstrate the flexibility of our approach in defining and deploying satellite network slices.


\bibliographystyle{IEEEtran}
\bibliography{IEEEabrv, final}

\end{document}